\documentclass[submission,copyright,creativecommons]{eptcs}
\providecommand{\event}{FM\&MDD 2017} 
\usepackage{breakurl}             
\usepackage{underscore}           
\usepackage{amssymb}
\usepackage{graphicx}
\usepackage{listings}
\usepackage{rotating} 
\usepackage{lscape}
\usepackage {bsymb,b2latex}
\usepackage[normalem]{ulem}
\usepackage{amsmath}
\usepackage{enumitem}
\usepackage{url}
\title{Incremental Database Design using UML-B and Event-B}
\author{Ahmed Al-Brashdi
\institute{University of Southampton\\ Southampton, UK}
\email{azab1g14@ecs.soton.ac.uk}
\and
Michael Butler
\institute{University of Southampton\\ Southampton, UK}
\email{mjb@ecs.soton.ac.uk}
\and
Abdolbaghi Rezazadeh
\institute{University of Southampton\\ Southampton, UK}
\email{ra3@ecs.soton.ac.uk}
}
\def\titlerunning{Incremental Database Design}
\def\authorrunning{A. Al-Brashdi, M. Butler \& A. Rezazadeh}
\begin{document}
\maketitle

\begin{abstract}
Correct operation of many critical systems is dependent on the data consistency and integrity properties of underlying databases. Therefore, a verifiable and rigorous database design process is highly desirable. This research aims to investigate and deliver a comprehensive and practical approach for modelling databases in formal methods through layered refinements. The methodology is being guided by a number of case studies, using abstraction and refinement in UML-B and verification with the Rodin tool. UML-B is a graphical representation of the Event-B formalism and the Rodin tool supports verification for Event-B and UML-B. Our method guides developers to model relational databases in UML-B through layered refinement and to specify the necessary constraints and operations on the database.
\end{abstract}

\section{Introduction}
Database systems hold large resources upon which critical decisions rely. These resources and decisions can be part of safety or business critical domains like health and patient systems or enterprise intelligent systems. This emphasises the fact that database systems are a very important field in software engineering~\cite{connollydatabase} and thus require a verifiable and rigorous design and implementation. While the chances of inconsistency and ambiguity of specifications are low in small and simple systems, the chances increase as the database size and complexity grow. The conventional database design using Entity-Relational Diagram (ERD) to describe databases is restricted to modelling the structure of the databases without specifying the system behaviour. Modelling only the structure of the databases does not prove its consistency or unambiguity as these can be caused by an operation of the database.
\par
This research tries to address the question of how to gradually design databases and prove their consistency and integrity. To address this, we propose a method for model-based database design in formal methods using a UML-like notation that supports layered refinement of system models. We use Event-B which is a formal method for rigorous specification and verification of digital systems~\cite{abrial}. It has been supported in an open tool platform called Rodin ~\cite{rodin}. We model our system using a UML-like notation in the Rodin tool called UML-B~\cite{umlb} which supports modelling in class diagrams and state machine diagrams. The UML-B tool translates UML-B models to Event-B models and the Rodin tool is used to verify their consistency. As class diagrams are commonly used to model database systems, using UML-B class diagrams will be more straightforward for database designers than modelling databases directly in Event-B.

This paper is structured as follows: In Section \ref{background}, we give background about the topics that are related to this research, mainly formal methods and relational databases. Section \ref{modelling} describes how to model an information system using UML-B and Event-B in Rodin through abstraction and refinement following case studies. In Section \ref{tool}, we outline our tool, UB2DB, which automatically generates database code from UML-B and Event-B model. 
Before concluding in section \ref{conclusion}, we outline the most related work in Section \ref{related}.

\section{Background}\label{background}

\subsection{Event-B}
Event-B is a formal method for system modelling and verification. A model in Event-B consists of a static part in a \emph{context} that defines the types, constants and axioms, and a dynamic \emph{machine} component with all \emph{variables} and \emph{invariants} as well as \emph{events} that change the variable state. An event may have \emph{guards} that must hold before the execution of the event. Event \emph{actions} change the state of a variable. All events in a machine must preserve all of its invariants. 

Refinement in Event-B enables a modeller to gradually specify the system at different levels~\cite{abrial}. Model refinement is a key concept in Event-B which enables a modeller to gradually specify the system so it becomes more precise and closer to reality \cite{abrialrefinement}. In a refinement, we start with an abstract model that describes the main functionality of a system. Then gradually we elaborate the system by adding further details in the specifications. Event-B refinement can be \emph{horizontal} or \emph{vertical}. The horizontal refinement includes adding extra details to the model while the vertical refinement or \emph{data refinement} transforms the state of an abstract variable to make it closer to programming implementation. 
\par
Applying refinement to a context in Event-B can be done by adding new sets, constants and axioms. Machine refinement may include adding new variables, invariants and new events or refining existing events \cite{abrial}. A new event in a refinement refines a \emph{skip} event that does nothing to the abstract model. When doing refinement we have to make some proofs so that the new refinement doesn't violate the abstract model. Moreover, doing a data refinement that removes an abstract variable and replaces it with another more concrete one introduces the \emph{glueing invariant}. The glueing invariant links between an abstract state and a concrete one \cite{abrial}.

Compared to other formal notations and methods such as Z \cite{znotation}, VDM \cite{vdm} and B method \cite{abrialb}, Event-B provides more flexible refinements in which we can introduce new events in refinements. This feature is important in our research as databases include different operations on their data and we need to introduce these operations on data when their variables are introduced in refinements.

\subsection{UML-B}
UML-B is a graphical notation for formal modelling in Event-B that is based on UML ~\cite{umlb}. A tool, called iUML-B, is provided which supports building diagrams in UML-B and is integrated into an Event-B machine or context. The model is translated into Event-B for verification. UML-B supports modelling with class diagrams which are used to describe the data structure and behavioural part of the system \cite{umlb}, and state diagram which are attached to a class, partitions the class instances into different states.

UML-B allows the modeller to choose one of three kinds when adding an event to a class. These kinds are \emph{normal}, \emph{constructor} or \emph{destructor} events. A constructor event should be selected for events that aim to create an instance of a class. The destructor is used for the opposite. For other operations, the normal event is selected; it adds a guard automatically to check that the instance to select or update is an element of that class set.

A class diagram in UML-B can be refined by adding new attributes or new associations. New classes and events are also possible when refining a UML-B class diagram and new class invariants can be defined. A state machine can be refined by adding new states ans transitions. A refinement of a state machine can also include adding nested states inside another state.

\subsection{Relational Database}
In 1970, Codd \cite{codd1970} introduced the relational model of database systems, which became the most widely used database type. The relational model of a database is composed of several \emph{relations}. Relation elements are represented as \emph{tuples} in which they may be formed by one or many \emph{columns} where each column is a set. Given sets $S_1, S_2, \dots, S_n$, relation $R$ is a set of $n$-tuples where each tuple has its first element from $S_1$, its second element from $S_2$, \dots etc. Each column in a relation has a heading (name) and a domain of values such as character or integer. Relations are viewed as tables, tuples as rows and columns as attributes.
\par
Any committed state of a database must guarantee and preserve some pre-defined constraints and assertions. These constraints might relate to a table, an attribute in a table, or a relation between one table and another. It is important that database consistency is proved so that any requirement of the database system is not violated in any valid state of the system. It is also important to prove that database requirements don't contradict with each others. This can achieved by using formal methods in which invariants are used to model these requirements. The invariants are universally quantified in the formal model and preserved by all its operations.

\section{Modelling database through abstraction and refinement}\label{modelling}
This section shows how to structure a database model in UML-B using our approach of layered refinements. In order to illustrate our approach clearly, we need to introduce some concepts that we identified when modelling three case studies. These case studies concern a Student Enrollment and Registration System (SRES), a Car Sharing System and an Emergency Room System.

Following the modelling of the case studies, we can generalise guidelines for modelling information systems in layered refinements by extending each refinement with extra features and complexity. The guidelines define both the structure of the model as the operations on its variables. We define how to model CRUD (Create, Read, Update and Delete) operations in Event-B with minimal mathematical notations that can be later translated automatically to database code.

Modelling databases gradually using abstraction and refinement in Event-B should help the modeller breaks down the complexity of a system. Since the refinement proofs are generated when using refinement in Event-B, the modeller can make sure that the refinement does not violate the abstract model.

\subsection{Primary, Secondary and Attribute Classes}
Modelling these case studies introduces different class types that can be used in defining a refinement strategy when modelling database systems. These classes are \emph{primary}, \emph{secondary} and \emph{attribute} classes. Primary classes are the classes that describe the main entities of the system and can be seen in classes that describe people such as \emph{Student} and \emph{Staff} in SRES, \emph{Member} in Car Sharing, or \emph{Patient} and \emph{Doctor} in Emergency Room. Primary classes can also illustrate activities such as \emph{Module} and \emph{Treatment}, objects such as \emph{Car} and \emph{Room}, or an organisation such as \emph{Department}.

\emph{Secondary} classes are the classes that relate two or more primary classes together. Examples of such classes are \emph{Registration} of a student in a particular model, \emph{Booking} of a car by a member, and \emph{Admission} of a patient by a doctor into a room. 

\emph{Attribute} class is a class that represents a complex attribute of a primary class that consists of multiple attributes. An example of such a class is the \emph{Address} of a person which by itself has attributes such as street number and postcode. For each concept and refinement, this section will show it using an example of one of the case studies.

\subsection{Modelling classes and their associations}
\begin{figure}
\centering
	\includegraphics[width=.5\linewidth]{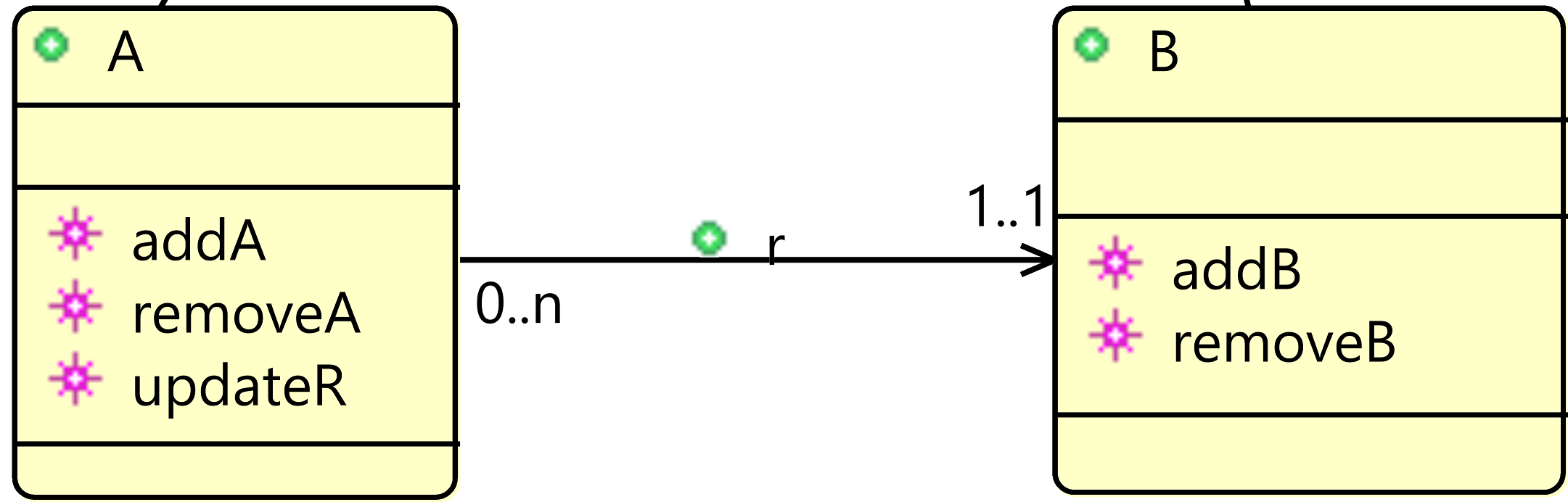}
	\caption{Abstract model of UML-B class diagram}
	\label{fig:m0model}
\end{figure}

Modelling information systems in our approach is done in different refinements that are defined in successive steps. The abstract model of the system may consist of different classes and the associations between them. 
Figure \ref{fig:m0model} shows an abstract UML-B model of two classes \emph{A} and \emph{B}. The diagram shows the classes and association between them, \emph{r}. 

\begin{figure}
	\includegraphics[width=\linewidth]{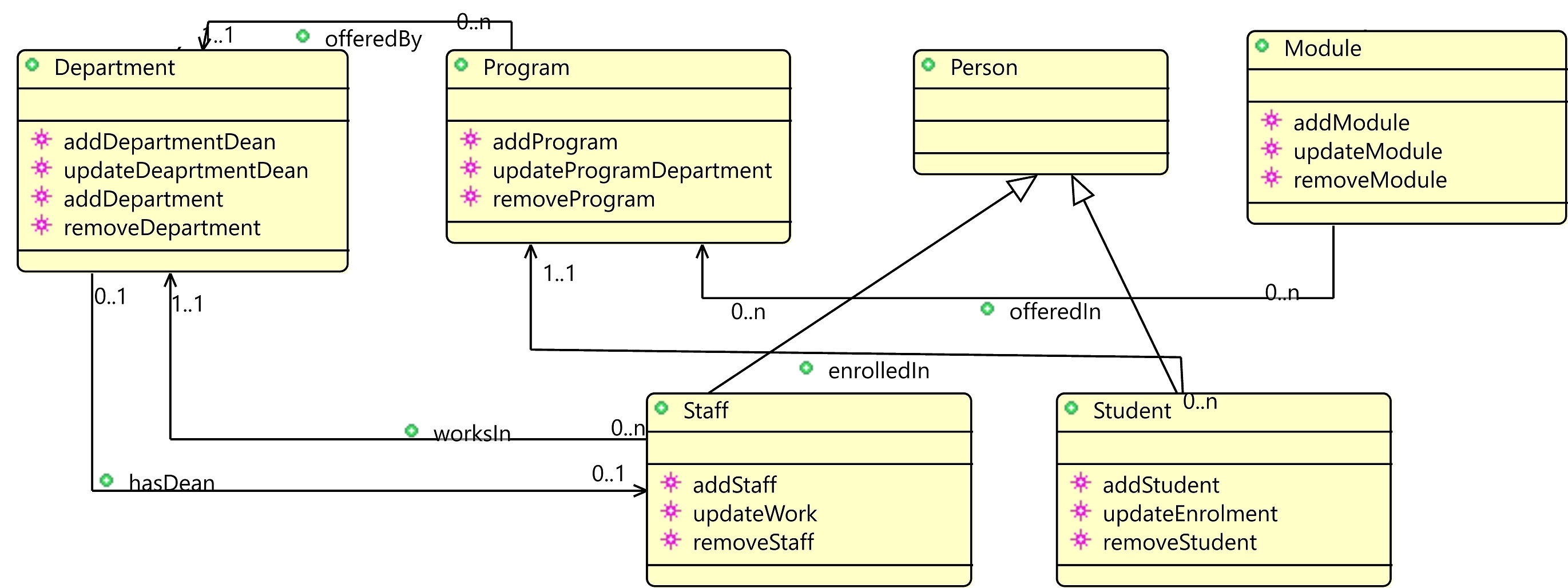}
	\caption{Abstract model of SRES entities and relations as a UML-B class diagram}
	\label{fig:m0}
\end{figure}

The abstract model needs to include all required events that modify the state of variables introduced in that model. Such events are \emph{add}, \emph{update} and \emph{remove} events, where add events insert new elements in the class, update events change the value of one or more of its attributes and remove events delete one or more records from it.  
Examples of primary classes in the SRES case study are \emph{Student}, \emph{Module}, \emph{Staff}, \emph{Program} and \emph{Department}. Figure \ref{fig:m0} shows our abstract UML-B model of the SRES case study.
\par
The model includes all events that change the state of its classes instances, attributes and associations. Add events are set as constructor event types in UML-B. For each class such as \emph{Program}, all associations from it to another class are added in its constructor event. For example, \emph{addProgram} for the \emph{Program} class has a parameter for \emph{offeredBy} and an action to map it to \emph{Program} instance as in action \emph{act2}.

\begin{center}
\EVT {addProgram}\cmt{ }
		\begin{description}
		\AnyPrm
			\begin{description}
			\ItemY{this\_Program,\ d }{ }
			\end{description}
		\WhereGrd
			\begin{description}
			\nItemY{ grd1 }{ this\_Program \notin  Program }{ } 
			\nItemY{ grd2 }{ d \in  Department,\ this\_Program \in  PROGRAM }{ } 
			\end{description}
		\ThenAct
			\begin{description}
			\nItemY{ act1 }{ Program :=  Program \bunion  \{ this\_Program\}  }{  }
			\nItemY{ act2 }{ \textit{offeredBy} :=  \textit{offeredBy} \bunion  \{ this\_Program\mapsto d\}  }{  }
			\end{description}
		\EndAct
		\end{description}
\end{center}
The model also includes inheritance between super and sub-classes. An example is the \emph{Staff} and \emph{Student} classes which are sub-classes of \emph{Person} class. The subclasses could have some explicit associations for each that are not shared between them.

Modelling the relation between \emph{Staff} and \emph{Department} introduces a \emph{circular dependency} in which each class relates to the other one forming a circle as a \emph{Department} has a dean and a member of \emph{Staff} works in a \emph{Department}. By modelling this in Event-B and specifying each association as a total function, both adding \emph{Staff} and adding \emph{Department} events are not enabled as each requires an instance from the other class. To avoid this, we weakened one association, \emph{hasDean}, by making the association optional, or partial function.

\subsection{Adding attributes and extending events}
\begin{figure}
\centering
	\includegraphics[width=.7\linewidth]{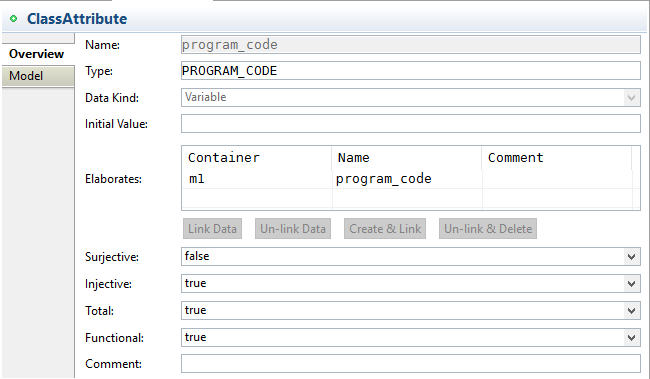}
	\caption{Setting class attribute in UML-B}
	\label{fig:attType}
\end{figure}

In a refinement, each class will have attributes that add some details about the class such as \emph{program\_code} in class \emph{Program}. After defining these classes and their associations, we refine the model by adding different attributes to each class and defining their constraints. The constraints such as \emph{not null} and \emph{unique} constraints can be defined by defining the attribute as \emph{total} and \emph{injective} functions when added in UML-B as in Figure \ref{fig:attType} for the attribute \emph{program\_code} in \emph{Program} class. Adding this as a refinement is because we prefer to have the general structure of the classes and associations between them first, then to add details to each individual class.

\begin{figure}
\centering
	\includegraphics[width=.7\linewidth]{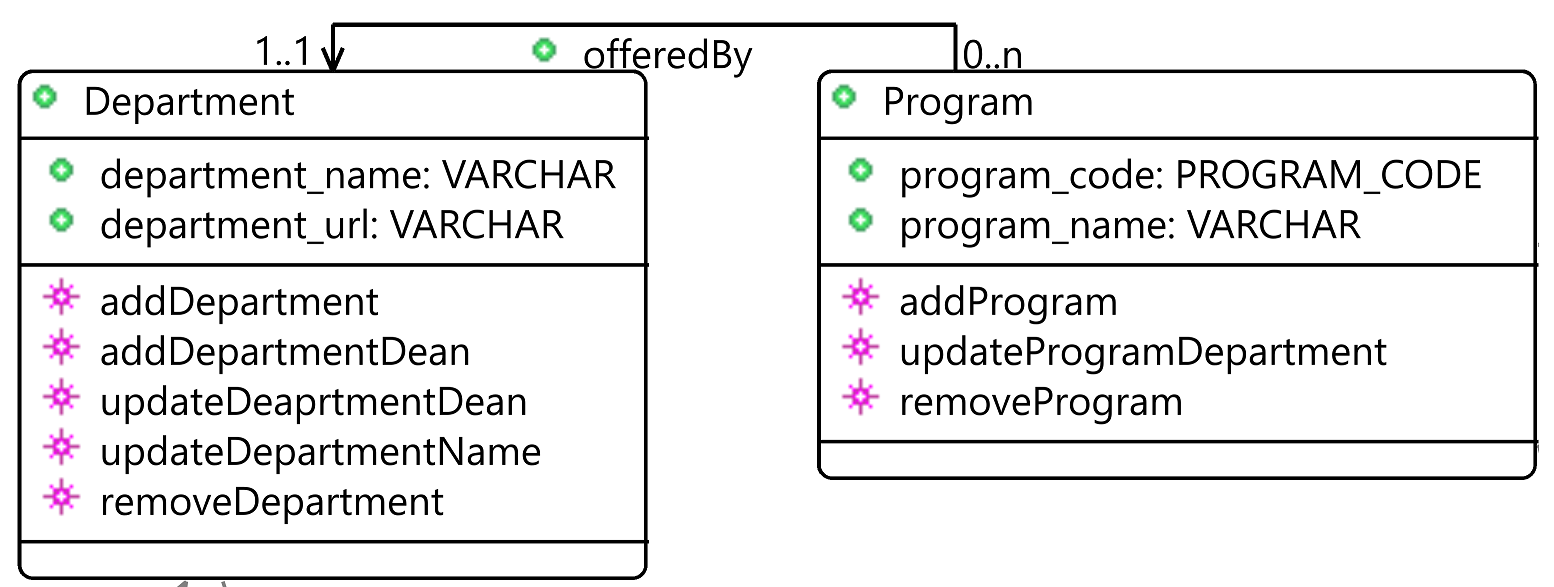}
	\caption{Adding attributes to the main classes}
	\label{fig:m1}
\end{figure}

Figure  \ref{fig:m1} shows this refinement in our approach where we added the attributes to the classes.  In this level, data types such as date and variable characters are defined as carrier sets and used as types for different attributes in various tables. All events are extended to include the new attributes such as \emph{program\_code} and \emph{progran\_name} in \emph{addProgram} event:
\begin{center}
    \EVT{addProgram}\cmt{\\ }
	\EXTD {addProgram}
		\begin{description}
		\AnyPrm
			\begin{description}
			\ItemY{p\_code,\ p\_name}{}
			\end{description}
		\WhereGrd
			\begin{description}
			\nItemY{grd4}{p\_code \in{} PROGRAM\_CODE,\ p\_code \in{} PROGRAM\_CODE}{\\}
			\end{description}
		\ThenAct
			\begin{description}
			\nItemY{act3}{program\_code \bcmeq{} program\_code \bunion{} \{this\_Program\mapsto{}p\_code\}}{\\}
			\nItemY{act4}{program\_name \bcmeq{} program\_name \bunion{} \{this\_Program\mapsto{}p\_name\}}{\\}
			\end{description}
		\EndAct
		\end{description}
\end{center}

\subsection{Modelling secondary classes}
\begin{figure}[h!]
\centering
	\includegraphics[width=.7\linewidth]{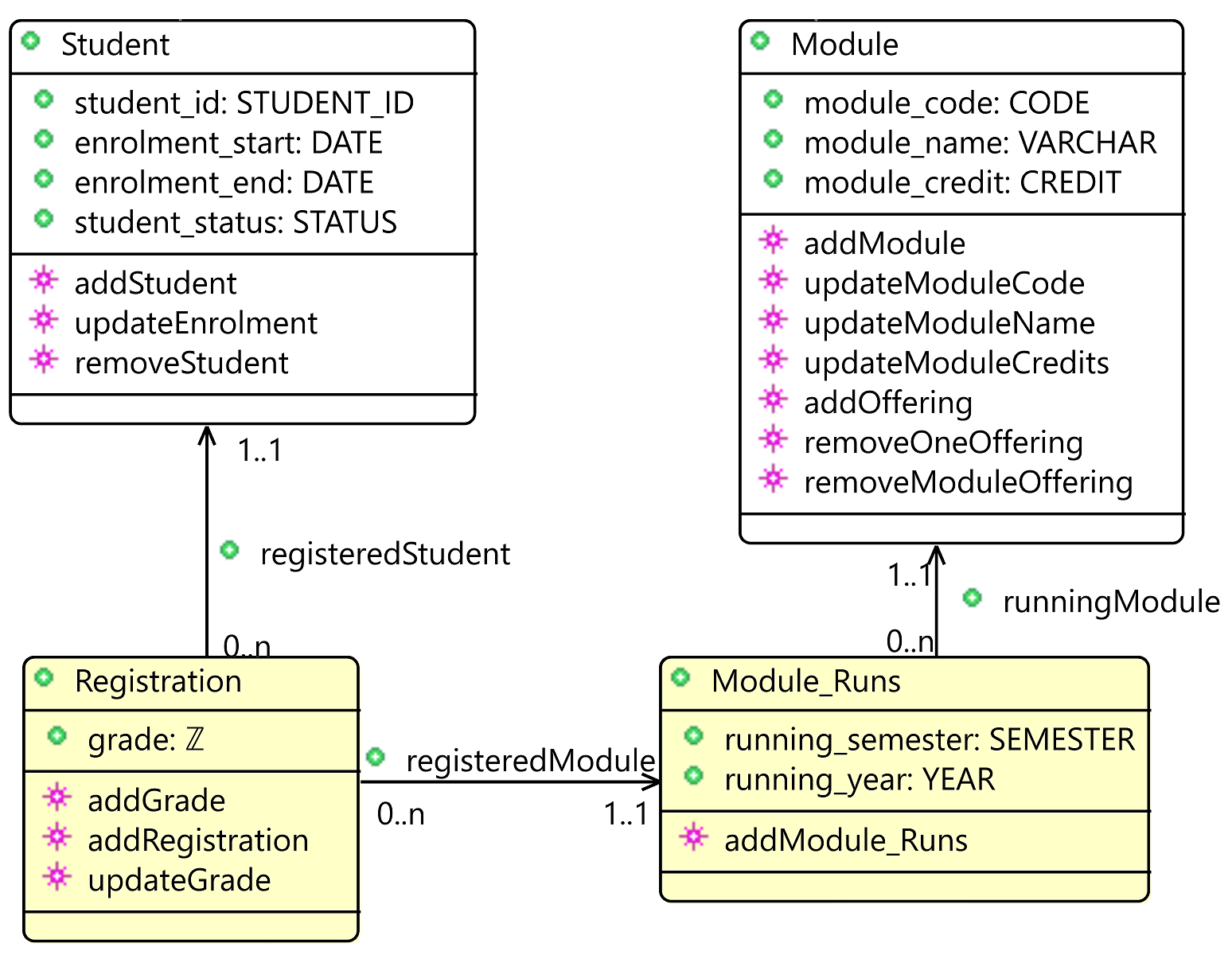}
	\caption{Adding secondary classes}
	\label{fig:m2}
\end{figure}

In a further refinement we introduce the secondary classes to the model in which they associate between primary classes or are instances of a primary class such as the \emph{Registration} in Figure \ref{fig:m2} which is a class that describes a Student taking a Module in a specific time and the \emph{Module\_Runs} which specifies modules running at given year and semester.

\subsection{Modelling attribute classes}
\begin{figure}[h!]
\centering
	\includegraphics[width=.7\linewidth]{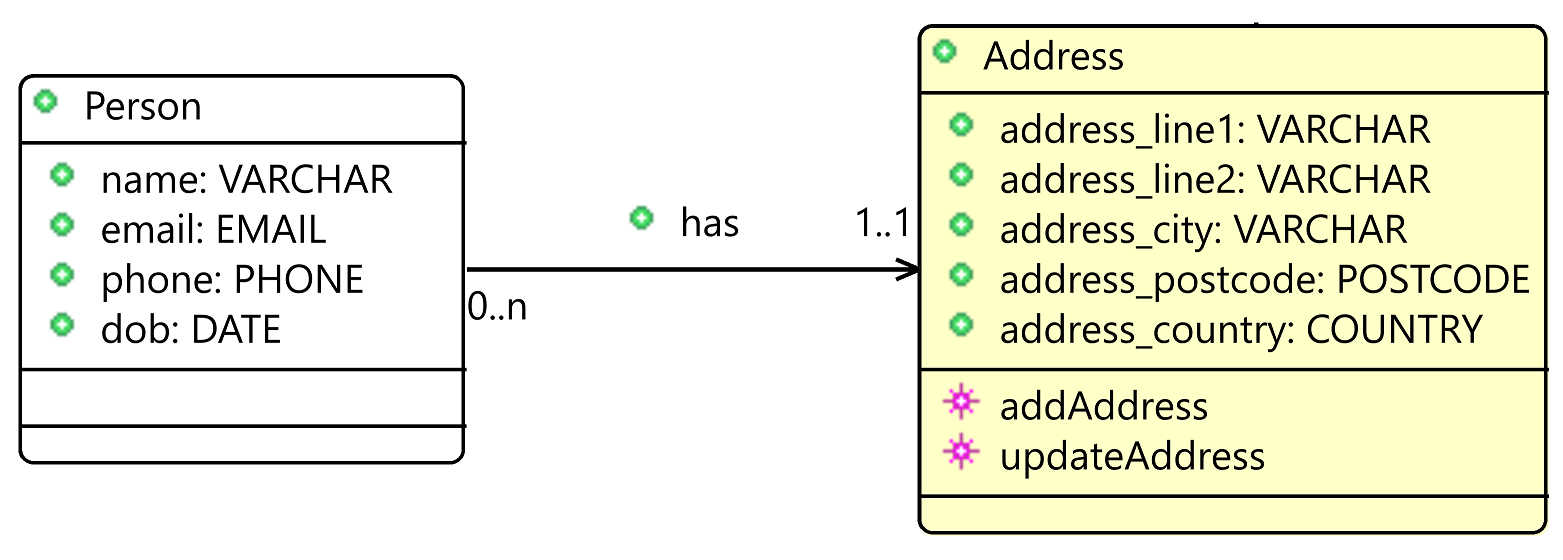}
	\caption{An example of an outer entity}
	\label{fig:m3}
\end{figure}

Another distinction is introduced in this model: the attribute classes. An example is \emph{Address} class, which is an attribute type that is associated with \emph{Person} as shown in Figure \ref{fig:m3}. The association is directed to and not from the Person class giving the assumption that each person might have $0..n$ addresses. This concept, the attribute class, can be introduced in any refinement. The association is defined from the primary/secondary class to the attribute class.

\subsection{Modelling historical data}
\begin{figure}
\centering
	\includegraphics[width=.7\linewidth]{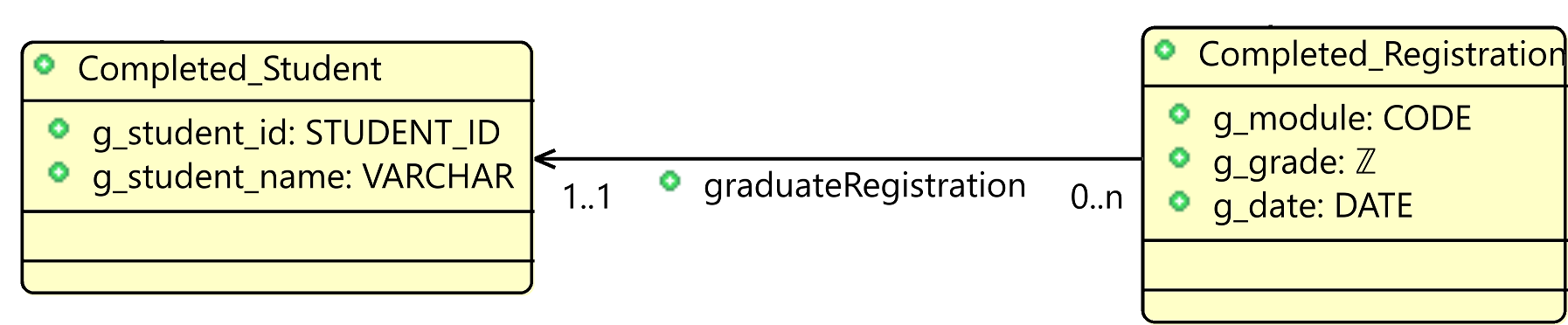}
	\caption{Historical data classes}
	\label{fig:m4}
\end{figure}

In some systems such as SRES, there might be a need to move some historical data into different tables from which it can be retrieved later. Moving the data is necessary when the table becomes large and a full scan becomes very expensive. For example, after a student has completed and passes his/her registered modules, instead of keeping all the records in the original or live table, the completed records will be moved into a historical table. While the new table is a subtype of the same supertype as the live table, they are not bound to the live one. \emph{Completed\_Student} and \emph{Completed\_Registration} in Figure \ref{fig:m4} are new classes that represent archives of the records of completed students. When a student finishes his or her degree, the information is moved to \emph{Completed\_Student} and the history of the registration is moved to \emph{Completed\_Registration}. We remove an instance from the \emph{live} class and add it to the \emph{historical} one in one event which is atomic in Event-B. The historical classes might have new attributes that are not in the live classes such as \emph{d\_date} which specifies the date of completion. This refinement can be introduced later in the system as it concerns archiving old data in which the modeller do not need to worry about it when the modelling starts. Similar refinement could include classes that are used to track the changes in the database in which it keeps logs of all the changed data and by whom.

\subsection{Association splitting} \label{splitting}
\begin{figure}[ht]
\centering
	\includegraphics[width=0.6\linewidth]{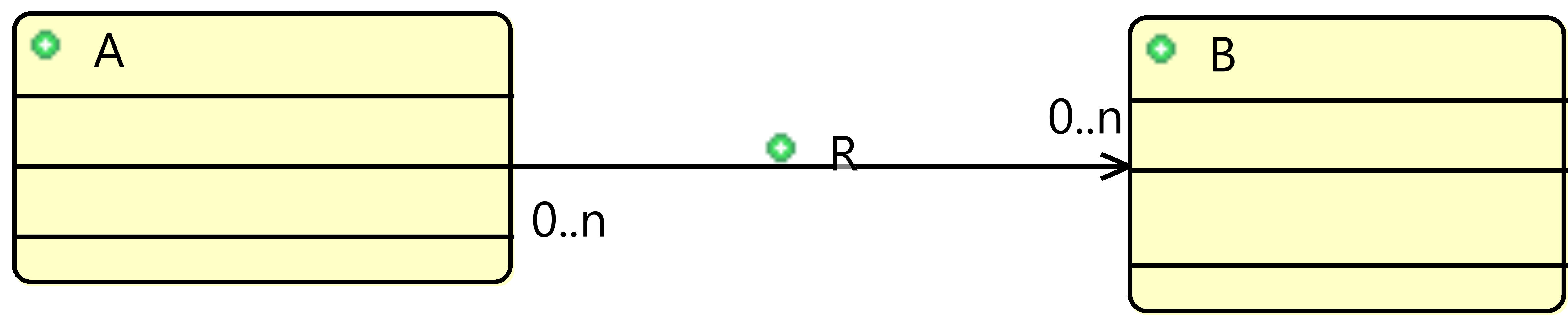}
	\caption{The abstract model of a relation R}
	\label{fig:relation0}
\end{figure}

\begin{figure}[ht]
\centering
	\includegraphics[width=0.7\linewidth]{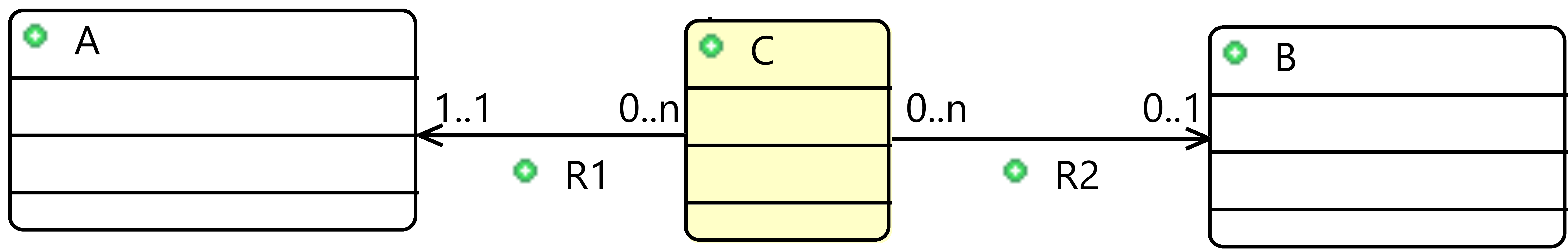}
	\caption{The refinement of relation R to R1 and R2}
	\label{fig:relation1}
\end{figure}

While association between two classes in UML-B can be of a type relation which is a many-to-many association, relational database model does not support direct many-to-many relationships. We need association splitting to make the formal specification closer to the implementation. For any relation in Event-B that is a many-to-many association between two classes, we introduce a design pattern, \emph{association splitting}, by refining it into a new class with two functions to the other two classes. This pattern as in Figures \ref{fig:relation0} and \ref{fig:relation1} shows the refinement of relation $R$ to two functions $R1$ and $R2$ from a newly created intermediate class $C$ to $A$ and $B$. The following gluing invariant, \emph{inv1}, specifies that $R$ is equal to the relational composition of inverse $R1$ ($R1$\textasciitilde) and $R2$: 

\begin{figure}[ht]
	\includegraphics[width=0.2\linewidth]{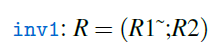}
\end{figure}
Since a relation does not have a duplication in pairs, the refined functions $R1$ and $R2$ must satisfy the same uniqueness of R as in \emph{inv2}. 
\begin{center}
    \begin{description}
        \nItem{inv2} {\forall a,b,c1,c2\qdot c1\mapsto a \in R1 \land c2\mapsto a  \in R1 \land \\                 \-\hspace{2.7cm}c1\mapsto b \in R2 \land c2\mapsto b \in R2 \limp c1=c2)}
    \end{description}
\end{center}
The second invariant specifies that we cannot have two \emph{C}s that both refer to the same \emph{a} and \emph{b}. This forms a \emph{composite} uniqueness in which the uniqueness is not about a single value, but the combination of multiple values. 

\subsection{Modelling Operations}
UML-B model provides three kinds of events: \emph{constructor}, \emph{destructor} and \emph{normal}. Our method and tool try to map these events to procedures that perform CRUD operations on the database. The guards in Event-B events must hold for an event to be enabled and must satisfy the model invariants.
\par
For a constructor event in UML-B, an instance of the class is created along with its attributes and associations such as \emph{addprogram}. A destructor removes an instance of a class with all its attributes and associations as in \emph{removeA} using domain subtraction. Domain subtraction $\{this\_a\} \domsub x$ removes all pairs of $x$ whose domain value is $this\_a$.
\begin{center}
    \EVT {removeA}\cmt{ }
		\begin{description}
		\AnyPrm
			\begin{description}
			\ItemY{this\_a }{ }
			\end{description}
		\WhereGrd
			\begin{description}
			\nItemY{ grd1 }{ this\_a \in  a }{ } 
			\end{description}
		\ThenAct
			\begin{description}
			\nItemY{ act1 }{a := a \setminus \{this\_a\},\ x := \{this\_a\} \domsub x,\ r := \{this\_a\} \domsub r }{  }
			\end{description}
		\EndAct
		\end{description}
\end{center}

Normal events in UML-B can be used to update or override set elements as in \emph{updateA} which uses function override to update the association \emph{r}. Function override of $r$ means that the range value that the domain $this\_a$ is mapped to is updated to the value $new\_r$. Normal events can be used also to query information from the classes as in \emph{getA} which retrieve all \emph{a}'s whose \emph{x} value is \emph{z}.
\begin{figure}[h!]
	\includegraphics[width=\linewidth]{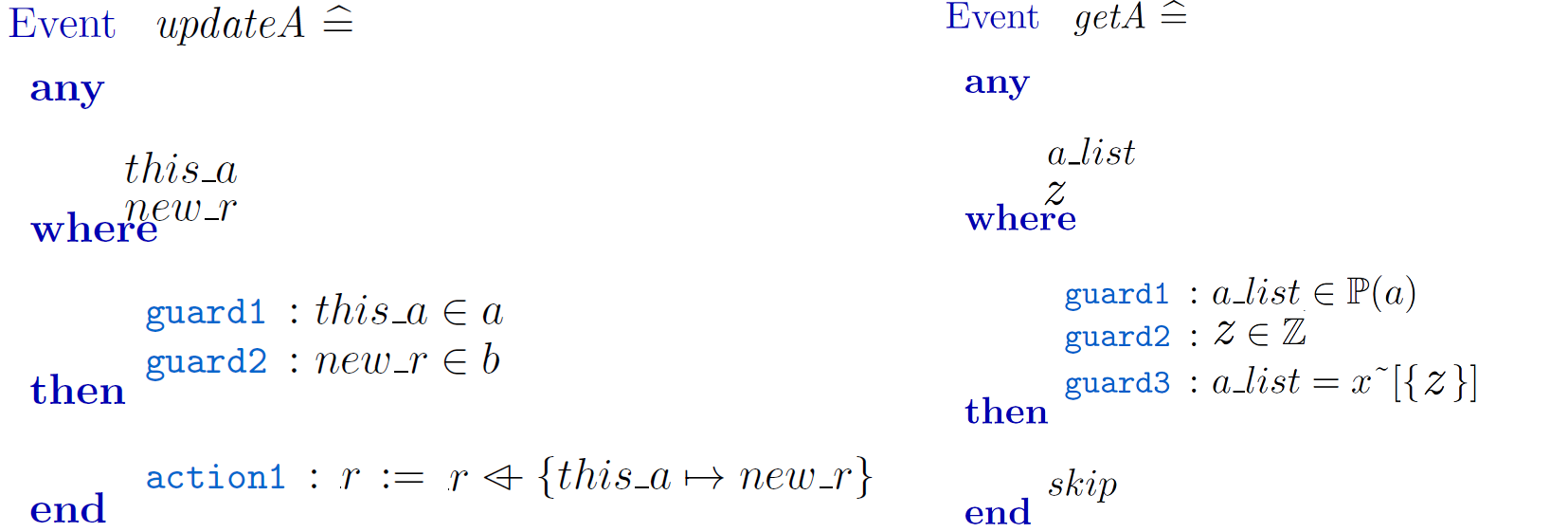}
	\label{fig:normalEvents}
\end{figure}

While the events that modify the state of machine variables are introduced in that machine, we introduced \emph{get} events in a later refinement because they might require a complete structure of different classes in order to retrieve valuable data. The event, \emph{getDepartmentStaff}, reports, in \emph{grd3}, all the Staffs working in a given department. The query, or get, events do not have actions as they just report some data from the model. 
\begin{center}
\EVT {getDepartmentStaff}\cmt{ }
		\begin{description}
		\AnyPrm
			\begin{description}
			\ItemY{d }{}
			\ItemY{staff\_list }{ }
			\end{description}
		\WhereGrd
			\begin{description}
			\nItemY{ grd1 }{ d \in  Department }{ } 
			\nItemY{ grd3 }{ \textit{staff\_list} \in  \pow(\textit{Staff}) }{ } 
			\nItemY{ grd3 }{  \textit{staff\_list} = worksIn^{-1} [\{ d\} ]}{ } 
			\end{description}
		\ThenAct
			\begin{description}
			\Item{ skip }
			\end{description}
		\EndAct
		\end{description}
\end{center}

\subsection{Summary of approach}
By using our approach to model an information system at different refinement levels, we introduced different concepts and distinctions in which each concept could be modelled in a new refinement. This approach can help the modellers to gradually model a complex system using layered refinement where in each refinement they focus on modelling and verifying a subset of the requirements. The approach can be summarised in the following steps:

\begin{itemize}
    \item Modelling primary classes,associations and relevant events.
    \item Introducing secondary classes, extending events and adding new events.
    \item Introducing attribute classes, extending events and adding new events.
    \item Introducing historical data, extending events and adding new events.
    \item Introducing query events.
\end{itemize}
The approach is extended further for extra features such as modelling database views. Two of our case studies with layered refinement can be found in \cite{casestudies}.

\subsection{Model verification}\label{verification}
\begin{figure}
\centering
	\includegraphics[width=.8\linewidth]{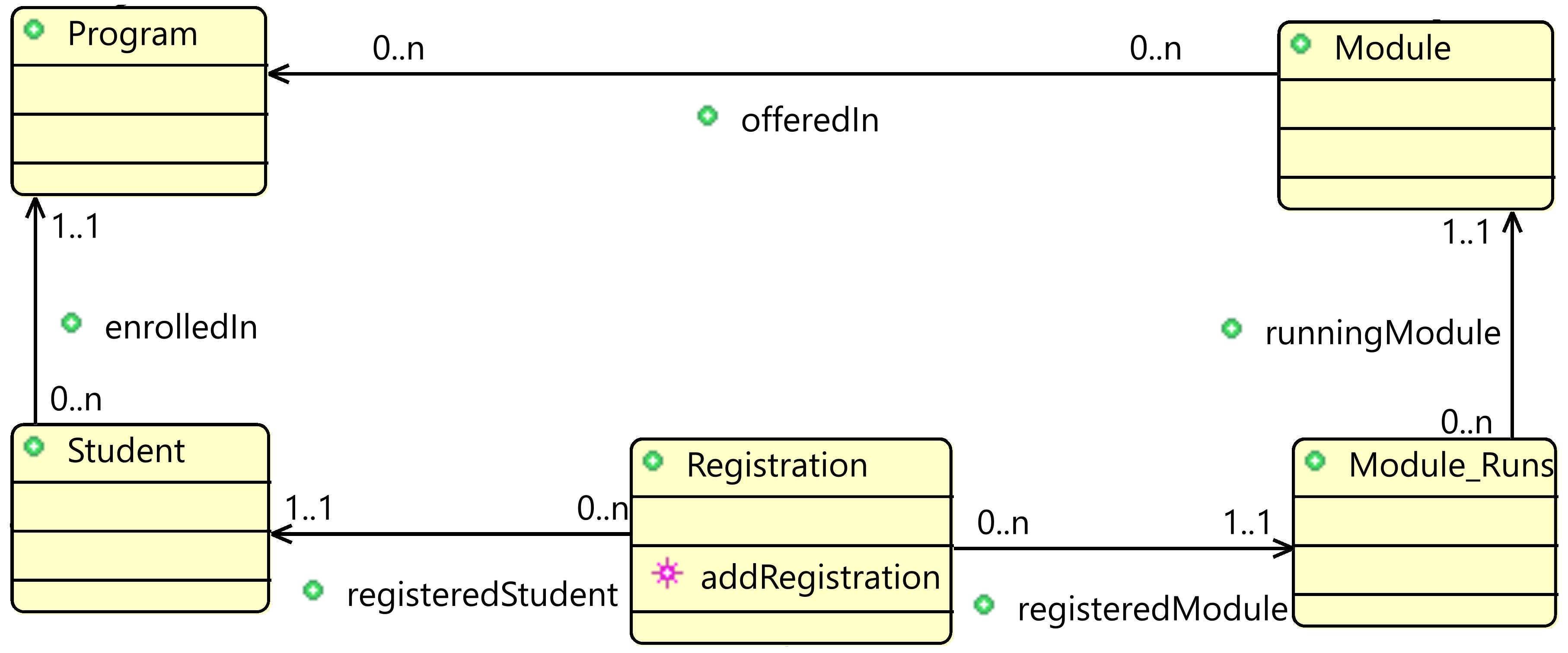}
	\caption{Student registration and enrollment}
	\label{fig:inv}
\end{figure}

By modelling databases in UML-B and Rodin, we introduce formal verification for our database models. The database constraints are modelled as \emph{invariants} in which they must be preserved by all events. Let's assume we have a requirement that Students can only register in Modules offered by the same program of study they are enrolled in as in Figure \ref{fig:inv}. This can be modelled by invariant \emph{inv1} which applies to all instances of the registration class. For presentation and space, \emph{inv1} is not shown in Figure \ref{fig:inv} and is added directly to Event-B machine. The invariant becomes universally quantified in Event-B:
\begin{center}
	\begin{description}
	\nItem{inv1}{\forall{}m,s\qdot{}s\mapsto{}m \in{} registeredStudent\converse{};registeredModule \limp{} \\ \-\hspace{2.1cm} runningModule(m)\mapsto{}enrolledIn(s) \in{} offeredIn}
	\end{description}
\end{center}
An event that adds a registration must preserve this invariant by having \emph{grd2} that ensures the module to register the student for is offered in the student program of study:
\begin{center}
    \EVT{addRegistration}\cmt{\\ }
		\begin{description}
		\AnyPrm
			\begin{description}
			\ItemY{this\_Registration,\ m,\ s}{}
			\end{description}
		\WhereGrd
			\begin{description}
			\nItemY{grd1}{this\_Registration \notin{} Registration,\ m \in{} Module\_Runs,\ s \in{} Student}{\\}
			\nItemY{grd2}{runningModule(m) \mapsto{} enrolledIn(s)\in{}offeredIn}{\\}
			\end{description}
		\ThenAct
			\begin{description}
			\Item{...}
			\end{description}
		\EndAct
		\end{description}
\end{center}

Without proving the consistency of this requirement among all operations on the database, a Students can register in a Module that is not offered by program of study in which they are enrolled in. The same verification applies to every invariant in the model in which it must be satisfied by every event in the model and any model that refines it. This draws an example for the importance of applying formal verification in the database design.  
\section{Tool support}\label{tool}

We developed a tool, called UB2DB \cite{UB2DB}, as a plugin for the Roding platform which generates the SQL code for the database from the UML-B class diagram. The tool supports translating to different constraints in SQL such as primary key, foreign key, not null, unique and check constraints. The events in the model are translated to \emph{stored procedures} in which event guards are validated and actions are executed in the stored procedure. A stored procedure is a program unit that is stored and validated in the database. The generated SQL code by the tool satisfies the system requirements and constraints and is validated against them for the case studies. We have generated the SRES database from the system and have been able to execute it successfully. Evaluating the performance of inserting 10000 records using the generated code shows that our code performed around 21\% slower than a hand written code. Further evaluation and optimisation should improve the efficiency and performance of UB2DB.

\section{Related Work}\label{related}
There is existing literature covering the concept of formalising database specifications. Schlatte and Aicherig present a database development of an industrial project using VDM-SL~\cite{dbvdm}. 
In~\cite{barros3} and \cite{khalafinejad}, the authors formalise relational databases in Z specifications. Barros in \cite{barros3} covers different CRUD operations as well as transactions, sorting, aggregations and other database components. There is no tool provided in which modellers can use to automatically generates database code for the formal definitions. Davies et al. in \cite{booster2odb} shows how to formalise an object-oriented databases using UML and Object Constraint Language (OCL) \cite{OCL} using Booster notation ~\cite{booster}. Mammar and Laleau in \cite{mammar1} have also specified relational database notions using UML-like notation. Their work supports modellers in designing databases using a UML diagram and then translate that model to a B specification and on to Java and SQL code. The refinement process supported by Laleau and Mammar work is toward a database implementation of B specifications. From Event-B, Wang and Wahls in \cite{eventb2sql} developed a Rodin plug-in that generates Java and JDBC code to create and query databases. However, the results shown issues with preserving database integrity from the code generated by their tool as in \cite{mymsc}.

None of these research provides general guidelines for modelling relational database in formal methods. Moreover, they do not address layered refinement where in each refinement a modeller can introduce new classes, attributes, associations and operations. The approach of modelling database systems by gradual refinement steps is an important aspect and contribution of this research. While layered refinement is well used when modelling in Event-B as stated is \cite{butler} and \cite{abrial}, the contribution of this research is applying that to database design with distinction of different concepts such as \emph{primary} and \emph{secondary} classes and different events for different database operations.
\section{Conclusion}\label{conclusion}
Formal modelling and specification of database systems is an important concept which has been covered in much literature. The importance of verified database design lies in critical domains and decisions that depend on correct and consistent data. The reviewed literature does not tackle how to structure the model in different refinement levels where in each refinement the modeller introduces some concepts for specification and verification. Where the refinement is used in the reviewed work, such as in \cite{mammar1}, it was a refinement for implementation where the concrete model becomes closer to an implementation language. Our research provides a practical approach for modelling the databases with different constraints through layered refinement. Throughout the process of modelling the case studies, we have identified the differences between different kinds of classes and events. These distinctions identify a refinement strategy or patterns for the model such as starting with classes and associations, then introduce attributes, then queries, etc. Undertaking the approach of specifying various components in different refinements enabled us to model each concept separately and verify its specifications. 
Our tool, UB2DB, which is developed to support our approach is validated against case studies and successfully generates the SQL and stored procedures code for database structure and operations from the case studies. While the presented approach for layered refinement are for guidance only, it is open for further extensions. Further extension patterns will be investigated using more and larger case studies. We will investigate different design patterns for modelling relational database which can be derived from different diverse case studies. The patterns will define how provide a solution for common problems when modelling information systems using UML-B and Event-B.

\nocite{*}
\bibliographystyle{eptcs}
\bibliography{references}
\end{document}